\documentclass[conference]{IEEEtran}

\IEEEoverridecommandlockouts

\usepackage{cite}
\usepackage{amsmath,amssymb,amsfonts}
\usepackage{algorithm}
\usepackage{algpseudocode}
\usepackage{graphicx}
\usepackage{multirow}
\usepackage{booktabs}
\usepackage{bbm}
\usepackage{subcaption}   
\usepackage{float} 

\DeclareFontFamily{U}{stix2bb}{}
\DeclareFontShape{U}{stix2bb}{m}{n} {<-> stix2-mathbb}{}

\NewDocumentCommand{\indicator}{}{\text{\usefont{U}{stix2bb}{m}{n}1}}

\begin{document}

\title{Diffusion-Based Generative Priors for Efficient Beam Alignment in Directional Networks}

\author{
\IEEEauthorblockN{
Esraa Fahmy Othman,~(Member, IEEE)\IEEEauthorrefmark{1}, 
Lina Bariah,~(Senior Member, IEEE)\IEEEauthorrefmark{1},
Merouane Debbah,~(Fellow, IEEE)\IEEEauthorrefmark{1}
}

\IEEEauthorblockA{
\IEEEauthorrefmark{1}\textit{Research Institute for Digital Future}, 
\textit{Khalifa University}, 127788 Abu Dhabi, UAE \\
Emails: 100062654@ku.ac.ae, lina.bariah@ku.ac.ae, merouane.debbah@ku.ac.ae
}
}

\maketitle

\begin{abstract}
Beam alignment is a key challenge in directional mmWave and THz systems, where narrow beams require accurate yet low-overhead training. Existing learning-based approaches typically predict a single beam and do not quantify uncertainty, limiting adaptive beam sweeping. We recast beam alignment as a generative task and propose a conditional diffusion model that learns a probabilistic beam prior from compact geometric and multipath features. The learned priors guide top-$k$ sweeps and capture the SNR loss induced by limited probing. Using a ray-traced DeepMIMO scenario with an 8-beam DFT codebook, our best conditional diffusion model achieves strong ranking performance (Hit@1 $\approx 0.61$, Hit@3 $\approx 0.90$, Hit@5 $\approx 0.97$) while preserving SNR at small sweep budgets. Compared with a deterministic classifier baseline, diffusion improves Hit@1 by about 180\%. Results further highlight the importance of informative conditioning and the ability of diffusion sampling to flexibly trade accuracy for computational efficiency. The proposed diffusion framework achieves substantial improvements in small-$k$ Hit rates, translating into reduced beam training overhead and enabling low-latency, energy-efficient beam alignment for mmWave and THz systems while preserving received SNR.

\end{abstract}


\section{Introduction}

Directional mmWave/THz communication requires accurate beam alignment to overcome severe path loss and blockage, but identifying the optimal beam from a large codebook during initial access introduces substantial latency. Recent work shows that site-specific deep learning (SSDL) can leverage ray-traced or measurement-based data to predict promising beams and reduce sweeping overhead~\cite{Heng2024SSBA,Heng2022GridFree}. Surveys of mmWave and THz beam management~\cite{BeamAlignSLR2025,BeamAlignV2X2024} highlight this transition toward data-driven beam training and emphasize the continued difficulty of maintaining reliability under non-line-of-sight (NLOS) and geometrically ambiguous conditions.

Deep learning methods for beam alignment typically operate as \emph{discriminative} models that output a single best beam, often without quantifying uncertainty. This limits their use in adaptive top-$k$ probing and makes them sensitive to errors in side-information. Pilot-efficient approaches~\cite{PilotOnlyBA2024} further reduce sensing overhead but retain the same determinism: they identify a single beam, not a distribution that reflects ambiguity or supports reliability--latency trade-offs. \\

Reinforcement learning has also been applied to beam alignment by learning online beam-sweeping policies \cite{che2025efficient}, but such approaches do not explicitly model beam uncertainty or provide a probabilistic beam distribution.

Generative diffusion models (GDMs) are probabilistic generative models that learn complex data distributions by reversing a gradual noising process. In the standard denoising diffusion probabilistic model (DDPM) formulation, data are corrupted through a forward diffusion process, and a neural network is trained to iteratively denoise samples back toward the data manifold \cite{nichol2021improved}. Deterministic diffusion implicit models (DDIM) provide an alternative sampling procedure that uses the same learned denoising model but follows a deterministic reverse trajectory, enabling faster or more controlled sampling while preserving generation quality \cite{song2020denoising}. 

Diffusion-based generative models have recently demonstrated strong performance across a range of wireless learning tasks, including sparse MIMO channel estimation~\cite{DiffusionCSI2024_WCL,Chen2025GDMVI}, RIS-assisted channel reconstruction~\cite{RISDiffusion2024}, location-conditioned channel synthesis~\cite{Lee2024Generating}, and precoder optimization in cell-free systems~\cite{zhang2025gdm}. These results indicate that diffusion models can effectively capture the structured variability of wireless propagation, making them well suited for uncertainty-aware and sample-efficient inference.
Wireless channels often contain multiple dominant propagation paths, rendering several beams simultaneously plausible. This motivates learning a probabilistic beam prior rather than a single deterministic prediction. Accordingly, we focus on discrete beam selection from compact side information to enable efficient and uncertainty-aware initial access.
Our main contributions are:
\begin{itemize}
    \item We cast beam alignment as a probabilistic inference problem and propose a conditional diffusion framework that learns uncertainty-aware beam priors.
    \item We design a lightweight conditioning diffusion mechanism based on compact geometric and propagation features, validated through targeted ablation studies.
    \item We show that diffusion-based priors substantially improve Hit@$k$ at small $k$ while preserving SNR, outperforming strong deterministic baselines and heuristic priors.
    \item We analyze DDPM and DDIM sampling strategies, revealing a clear performance--complexity trade-off for practical deployment.
\end{itemize}

\section{System Model}

We consider a single-cell downlink system in which a base station (BS) with a
uniform linear or planar array serves a user equipment (UE) at location
$x \in \mathbb{R}^d$. As in directional mmWave and THz systems, the BS selects a
transmit beam from a finite codebook during initial access, and beam sweeping is
used to identify a high-gain direction. The channel and beamforming structure
follow the ray-traced DeepMIMO ASU scenario \cite{alkhateeb2019deepmimo}, which
captures realistic geometry, angular statistics, and multipath propagation.

\subsection{Directional Channel Model}
The ray-traced data preserves the geometric and
angular structure relevant to directional beamforming. For each UE, the narrowband channel vector is modeled as a geometric multipath sum
\begin{equation}
    \mathbf{h} = \sum_{\ell=1}^{L} \alpha_{\ell}\,
    \mathbf{a}_{\mathrm{BS}}(\theta_{\ell}^{\mathrm{AoD}}),
    \label{eq:channel_model}
\end{equation}
where $\alpha_{\ell}$ and $\theta_{\ell}^{\mathrm{AoD}}$ denote the gain and departure angle of the $\ell$th path, respectively, and $\mathbf{a}_{\mathrm{BS}}(\cdot)$ is the transmit array response. The channel data provided in the ASU scenario correspond directly to~\eqref{eq:channel_model} through ray–traced multipath parameters (AoA/AoD, LOS/NLOS, path powers).

\subsection{Beamforming Codebook}
The BS employs a DFT-based beamforming codebook with $N_{\mathrm{beam}}$ beams,
\begin{equation}
    \mathbf{W} = [\,\mathbf{w}_1,\dots,\mathbf{w}_{N_{\mathrm{beam}}}\,] \in \mathbb{C}^{N_t \times N_{\mathrm{beam}}},
\end{equation}
where $N_t$ is the number of BS antennas and $\mathbf{w}_b$ denotes the $b$-th unit-norm beamforming vector. 
The beamforming gain of beam $b$ is
\begin{equation}
    g_b = \left|\mathbf{h}^{\mathrm{H}}\mathbf{w}_b\right|^2,
\end{equation}
where $\mathbf{h}$ is the narrowband BS--UE channel vector. 
Normalizing these gains yields a beam distribution
\begin{equation}
    p(b \mid x) =
    \frac{g_b}{\sum_{j=1}^{N_{\mathrm{beam}}} g_j},
\label{eq:beam_gain}
\end{equation}
which is computed using the true channel during training and serves as the supervision target for learning a feature-conditioned beam prior from the UE feature vector $x \in \mathbb{R}^d$. 
The optimal beam is $b^\star = \arg\max_b g_b$.

\subsection{Beam Sweep and Alignment Formulation}
\label{sec:beam_metrics}
During initial access, the BS is permitted to sweep a subset of beams. For a sweep budget $k$, the BS probes the $k$ beams with the largest predicted probabilities under a model $\hat{p}(b \mid x)$:
\begin{equation}
    \mathcal{S}_k = \mathrm{Top}\,k\big(\hat{p}(b\mid x)\big),
\end{equation}
where $\hat{p}(b\mid x)$ denotes the model predicted beam prior distribution.  
A successful sweep occurs if $b^\star \in \mathcal{S}_k$, defining
\begin{equation}
    \mathrm{Hit}@k = \indicator\!\left\{ b^\star \in \mathcal{S}_k \right\},
\end{equation}
where $\indicator\{\cdot\}$ is the indicator function. 

The link budget impact of beam prediction is quantified via the SNR ratio
\begin{equation}
    \label{eq:snr_ratio_single}
    \mathrm{SNR\,ratio}@1 =
    \frac{g_{\hat{b}}}{g_{b^\star}},
\end{equation}
where $\hat{b}=\arg\max_b \hat{p}(b\mid x)$ is the beam selected from the predicted prior and $b^\star$ is the optimal beam. We also evaluate a sweep-dependent SNR ratio
\begin{equation}
    \label{eq:snr_ratio_k}
    \mathrm{SNR\,ratio}@k =
    \frac{\max_{b \in \mathcal{S}_k} g_b}{g_{b^\star}},
\end{equation}
which compares the best gain among the swept beams to the gain of the optimal beam. In practice, SNR ratios are computed only for users whose optimal gain $g_{b^\star}$ exceeds a small floor (here $10^{-12}$), avoiding unstable ratios for links with vanishing power.
The learning objective is therefore to construct $\hat{p}_\theta(b \mid x)$, a well calibrated, uncertainty-aware beam prior conditioned on compact features $x \in \mathbb{R}^d$, such that Hit@$k$ is large and the SNR ratios in~\eqref{eq:snr_ratio_single}–\eqref{eq:snr_ratio_k} remain close to one.

\subsection{Dataset Description}

We evaluate the proposed framework using the ASU outdoor scenario from the DeepMIMO dataset.
Our experiments use a DFT codebook with
$N_{\mathrm{beam}} = 8$, producing an 8-dimensional beam-prior vector per UE.
From the scenario, we captured realistic geometry, multipath propagation, and angular diversity for a single BS serving a dense grid of UE locations. For each UE, the dataset provides ray-traced multipath parameters (AoA/AoD, path powers, delays, LOS/NLOS state) together with 3D coordinates $(x_{\mathrm{UE}}, y_{\mathrm{UE}}, z_{\mathrm{UE}})$. This setup serves as a proof-of-concept for learning probabilistic beam priors rather than a finalized system. The benefits are expected to be more pronounced for larger beam codebooks, as the model achieves high Hit@$k$ performance with very small sweep budgets ($k=1$–$5$).


\subsection{Beam Prior Construction}
For each UE, we compute the beam gains and normalized beam prior using the expressions in Section \ref{sec:beam_metrics}.

\subsection{Conditioning Features}

To assess the impact of side information on beam-prior learning, we construct
conditioning vectors with dimensionality $d \in \{3,5,7\}$, where $d$ denotes the
number of conditioning features. The feature sets progressively incorporate
geometric, propagation, and angular cues available at the UE.

\paragraph{3D geometric features}
This set contains only the UE location:
\begin{equation}
\mathbf{x}^{(3)} =
\begin{bmatrix}
x_{\mathrm{UE}},\, y_{\mathrm{UE}},\, z_{\mathrm{UE}}
\end{bmatrix}^{\mathsf{T}}.
\end{equation}

\paragraph{5D geometric and propagation features}
We augment the 3D coordinates with the BS–UE distance $d_{\mathrm{UE}}$ and a
binary LOS indicator:

\begin{equation}
\mathbf{x}^{(5)} =
\begin{bmatrix}
x_{\mathrm{UE}},\, y_{\mathrm{UE}},\, z_{\mathrm{UE}},\, d_{\mathrm{UE}},\, \indicator_{\mathrm{LOS}}
\end{bmatrix}^{\mathsf{T}}.
\end{equation}

$\indicator_{\mathrm{LOS}} \in \{0,1\}$, where 1 indicates LOS and 0 indicates NLOS.

\paragraph{7D geometric, propagation, and angular features}
Finally, we include normalized strongest-path AoA and AoD azimuths to capture dominant directional cues:

\begin{equation}
\mathbf{x}^{(7)} =
\begin{bmatrix}
x_{\mathrm{UE}},\,
y_{\mathrm{UE}} ,\,
z_{\mathrm{UE}} ,\,
d_{\mathrm{UE}} ,\,
\indicator_{\mathrm{LOS}} ,\,
\widehat{\theta}_{\mathrm{AoA}} ,\,
\widehat{\theta}_{\mathrm{AoD}}
\end{bmatrix}^{\mathsf{T}}.
\end{equation}

These feature groups enable controlled ablations on the value of geometric, propagation, and angular information for learning the beam prior in~\eqref{eq:beam_gain}.

\begin{figure*}[t]
    \centering
    \includegraphics[width=0.75\linewidth]{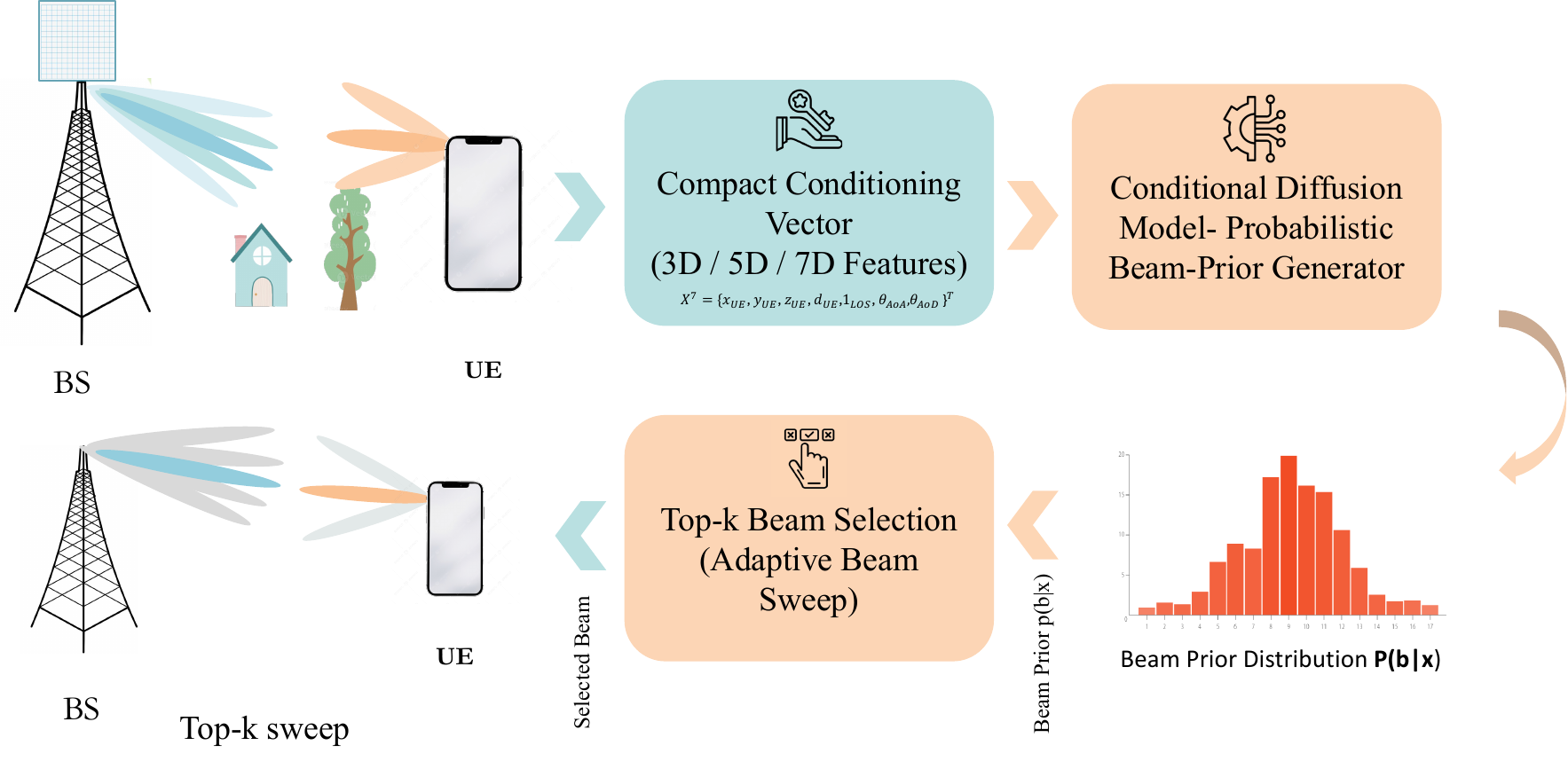}
    \caption{Overview of the proposed diffusion-based probabilistic beam alignment
framework. Compact geometric and multipath features are used to condition a
diffusion model that generates a beam-prior distribution, which then guides
adaptive top-$k$ beam sweeping for low-overhead alignment.}
    \label{fig:probabilistic_beam_alignment}
\end{figure*}
\section{Proposed Diffusion-Based Beam Prior Model}

The objective of the proposed framework is to train a conditional generative
model that maps lightweight side information $x \in \mathbb{R}^d$ to a
probability distribution over a discrete beam codebook, as illustrated in
Fig.~\ref{fig:probabilistic_beam_alignment}. Rather than predicting a single beam
index, the model generates a beam-prior vector that captures uncertainty in
directional propagation. These priors directly support adaptive top-$k$ beam
sweeps for efficient initial access.

\subsection{Diffusion Probabilistic Modeling}

Diffusion models learn a target distribution by progressively perturbing data
with Gaussian noise and training a neural network to invert this corruption
process. For a ground-truth beam prior $y_0$, the forward noising step at time $t$
is defined as
\begin{equation}
    q(y_t \mid y_0)
    = \mathcal{N}\!\left(
        \sqrt{\bar{\alpha}_t}\, y_0,\;
        (1 - \bar{\alpha}_t)\mathbf{I}
    \right),
    \label{eq:forward_diffusion}
\end{equation}
where $\bar{\alpha}_t$ is determined by the diffusion noise schedule. As $t$
increases, the noisy variable $y_t$ becomes increasingly dominated by Gaussian
noise.

The reverse denoising process is learned by a neural network $f_\theta(y_t, x, t)$ that predicts the noise injected at step $t$. Using this noise estimate, the reverse transition is constructed as
\begin{equation}
    p_\theta(y_{t-1} \mid y_t, x)
    = \mathcal{N}\!\big(\mu_\theta(y_t,x,t),\, \Sigma_t\big),
    \label{eq:reverse_process}
\end{equation}

where the mean $\mu_\theta(\cdot)$ is computed so as to invert the forward
corruption in~\eqref{eq:forward_diffusion}, and the covariance $\Sigma_t = \sigma_t^2 I$ is fixed and determined by the forward noise schedule, with $\sigma_t^2 = \beta_t \frac{1-\bar{\alpha}_{t-1}}{1-\bar{\alpha}_t}$, where $\beta_t$ denotes the predefined diffusion noise variance at step $t$. Conditioning is implemented by embedding the side information $x$ and concatenating it with the time embedding and the noisy beam representation before passing them into the denoiser. Iterating the learned reverse steps from $t=T$ to $0$ yields an estimate $\hat{y}_0$ of the clean beam-prior vector.

\subsection{Model Architecture and Variants}

The diffusion denoiser $f_\theta$ takes as input a sinusoidal embedding of the diffusion timestep $t$, the conditioning vector $x$, and the noisy beam vector $y_t$, which are jointly processed to predict the injected noise.

We evaluate two MLP-based denoisers that differ only in network capacity, with hidden dimensions of $256$ (small) and $512$ (large). Both share the same diffusion formulation, training objective, and conditioning mechanism.

In addition, we consider a UNet-based denoiser that replaces the simple fusion MLP with an encoder--decoder architecture with skip connections, while keeping the same conditioning features and diffusion schedule.

\subsection{Training Objective}

Following the DDPM formulation, the model is trained using a noise prediction
objective. For a randomly selected timestep $t$, a noisy sample is generated as
\begin{equation}
    y_t = \sqrt{\bar{\alpha}_t}\, y_0 +
          \sqrt{1 - \bar{\alpha}_t}\, \epsilon,
\end{equation}
where $\epsilon \sim \mathcal{N}(0,\mathbf{I})$. The denoiser produces a noise
estimate $\hat{\epsilon} = f_\theta(y_t, x, t)$ and is trained by minimizing
\begin{equation}
    \mathcal{L}(\theta)
    =
    \mathbb{E}_{y_0,\epsilon,t}
    \left[
        \|\epsilon - f_\theta(y_t, x, t)\|_2^2
    \right].
    \label{eq:mse_loss}
\end{equation}

A linear variance schedule is used during training
for all diffusion models.

\subsection{Sampling Procedure}
At inference time, the trained denoiser $f_\theta$ is fixed and used with different sampling procedures. Sampling starts from Gaussian noise $y_T \sim \mathcal{N}(\mathbf{0}, \mathbf{I}_{N_{\mathrm{beam}}})$ and follows the learned reverse diffusion process.


For DDPM sampling, all $T=500$ reverse steps are used with stochastic updates.
For DDIM sampling, a deterministic update rule is applied using only $50$ reverse
steps, substantially reducing inference latency
and energy while using the same learned denoiser.

Algorithm~\ref{alg:diffusion_beam} summarizes the generic diffusion-based
generation process.

\begin{algorithm}[t]
\caption{Generation of Beam Priors via Diffusion Sampling}
\label{alg:diffusion_beam}
\setlength{\baselineskip}{0.99\baselineskip}
\begin{algorithmic}[1]
\State Initialize $y_T \sim \mathcal{N}(0,I)$
\For{$t = T, \dots, 1$}
    \State Predict noise $\hat{\epsilon} = f_\theta(y_t, x, t)$
    \State Compute reverse mean $\mu_t(\hat{\epsilon})$
    \State Sample $y_{t-1}$ (DDPM) or apply a deterministic DDIM update
\EndFor
\State Normalize $\hat{y}_0$ to obtain the beam prior $p(b \mid x)$
\end{algorithmic}
\end{algorithm}

\section{Experimental Setup}

We evaluate the proposed diffusion-based beam prior model on the ASU ray-traced scenario described earlier. All models use the same $90\%/10\%$ train/validation split, performed randomly at the UE level such that each UE appears exclusively in either the training or validation set. All models are trained for 20 epochs using the AdamW optimizer with a learning rate of $10^{-3}$; diffusion denoisers use mini-batches of size 256 (small) or 512 (large), while baseline models use a batch size of 512. At inference, DDPM sampling uses $500$ reverse steps, while DDIM sampling uses $50$ deterministic reverse steps ($\eta=0$, where $\eta$ controls the amount of stochastic noise injected during DDIM sampling).
We compare diffusion against deterministic and generative baselines and perform ablations over conditioning dimensionality, denoiser architecture, denoiser capacity, and sampling strategy.

\subsection{Models Under Comparison}

We consider the following model classes.

\paragraph{Baselines}
We compare the proposed approach against several baseline methods, including a classifier (MLP) that predicts a single beam index, a regressor (MLP) that outputs a normalized beam score vector, and a variational autoencoder (VAE) that generates beam-prior distributions from the same conditioning features. We also consider heuristic and reference baselines, including an AoA-based method that selects the DFT beam closest to the strongest-path angle, as well as uniform and random priors.

\paragraph{Proposed diffusion models}
We evaluate conditional diffusion models with two denoiser capacities, \emph{small} (hidden dimension $256$) and \emph{large} (hidden dimension $512$), under three conditioning settings $d \in \{3,5,7\}$. Both DDPM and DDIM sampling procedures
are considered at inference.

\begin{figure*}[!htbp]
\centering
\subfloat[Hit@$k$ for diffusion and baselines]{%
    \includegraphics[width=0.44\linewidth]{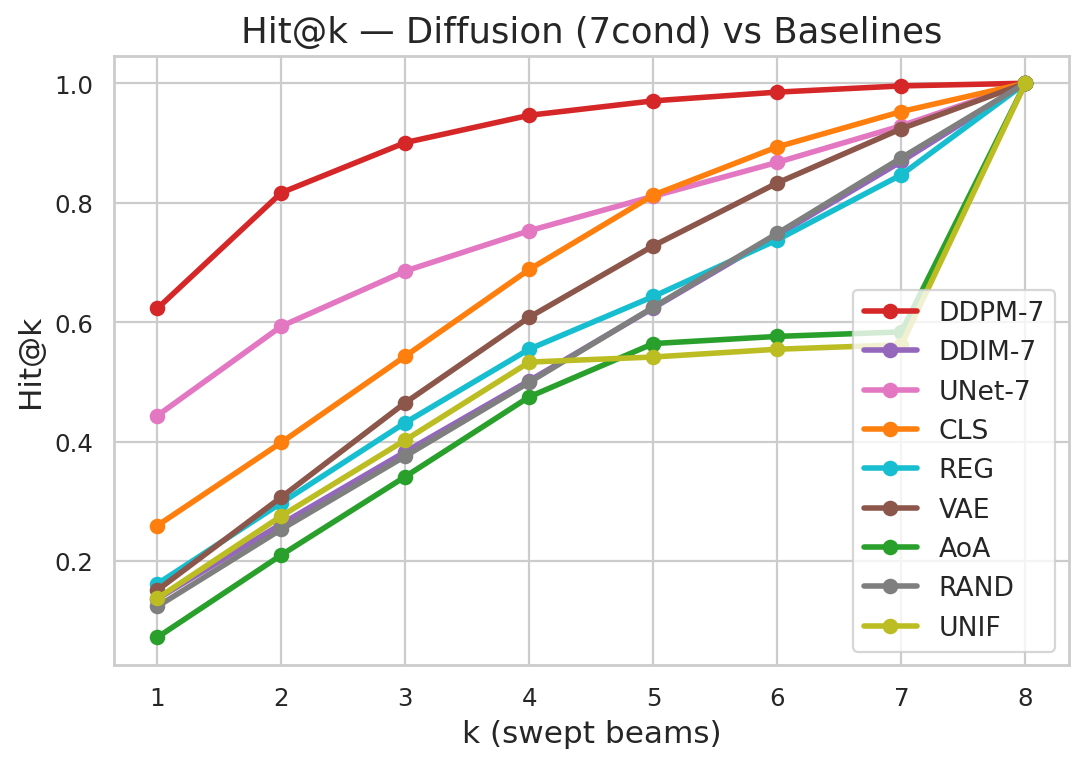}}
\hfill
\subfloat[SNR ratio for diffusion and baselines]{%
    \includegraphics[width=0.44\linewidth]{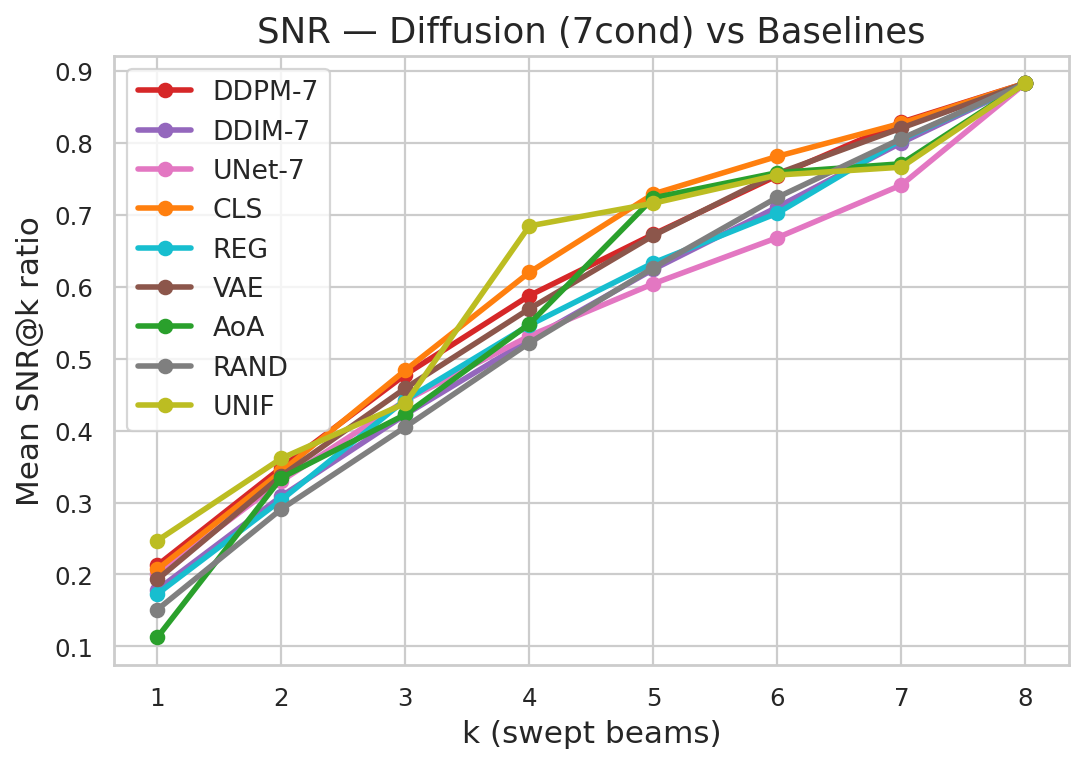}}
\vspace{0.15cm}

\caption{Baseline comparison of Hit@$k$ and SNR\,ratio@$k$. Probabilistic Diffusion (DDPM-7 and UNet-7) achieves substantially higher Hit@$k$ at small sweep budgets than deterministic and heuristic baselines with comparable SNR, while remaining competitive at larger $k$.}

\label{fig:baseline_results}
\end{figure*}

\subsection{Training Protocol}
Optimization is carried out using AdamW with matched hyperparameters across models. Classifier and regressor baselines are trained using cross-entropy and MSE losses, respectively, while the VAE employs a lightweight encoder–decoder architecture.

\subsection{Evaluation Metrics}

Beam alignment performance is evaluated using the metrics defined in Section~\ref{sec:beam_metrics}.

Hit@$k$ measures sweep efficiency by indicating whether the optimal beam is contained within the top-$k$ beams ranked according to the predicted prior distribution. SNR\,ratio@$k$ quantifies the link-budget loss resulting from limited beam sweeping and is computed only for users whose optimal beam gain exceeds a predefined threshold in order to avoid numerical instability. Together, Hit@$k$ and SNR\,ratio@$k$ provide a complementary characterization of sweeping overhead and received signal quality.

In addition, computational efficiency is assessed via inference latency and energy consumption. Latency is measured as the average per-user inference time on the validation set, averaged over 5 runs and including the full diffusion sampling loop. Energy consumption per user is measured on an NVIDIA A100 GPU by integrating GPU power readings over time using NVML and averaging over 3 runs.

\subsection{Diffusion Ablations}

We study four axes of variation that impact both beam alignment performance and inference cost.

\paragraph{Conditioning dimensionality}
Diffusion models are trained with conditioning dimensionality $d \in \{3,5,7\}$ to assess the role of geometric, propagation, and angular information in beam ranking.

\paragraph{Model capacity}
Both small (256-hidden) and large (512-hidden) denoisers are evaluated to determine whether performance is limited by network capacity or by the richness of side information.

\paragraph{Denoiser architecture}
We compare a baseline diffusion MLP denoiser with a UNet-based diffusion architecture that incorporates encoder--decoder structure and skip connections to better capture correlations in the beam domain, while sharing the same conditioning features and diffusion schedule.

\paragraph{Sampler type}
We compare DDPM (stochastic) and DDIM (deterministic) sampling strategies using the same trained denoisers. DDPM employs a larger number of denoising steps, while DDIM enables substantially reduced sampling depth at inference time, exposing a tradeoff between beam ranking accuracy and latency and energy consumption.

\section{Results and Analysis}

We evaluate the proposed diffusion-based beam prior model against deterministic (classifier, regressor), generative (VAE), and reference (AoA, uniform, random) baselines under a common dataset split and conditioning features. Performance is assessed using ranking accuracy and SNR preservation across sweep budgets.

\subsection{Baseline Comparison}

Figure~\ref{fig:baseline_results} compares the best-performing diffusion models with 7D conditioning (DDPM-7 and UNet-7) against learned and heuristic baselines in terms of Hit@$k$ and SNR\,ratio@$k$.

\paragraph{Diffusion vs.\ learned baselines}
At small sweep budgets, diffusion-based methods clearly outperform learned baselines, including the classifier (CLS), regressor (REG), and variational autoencoder (VAE). In particular, DDPM-7 achieves Hit@1 of approximately $0.61$, compared to about $0.22$ for CLS, $0.18$ for REG, and $0.17$ for VAE. As $k$ increases, CLS and VAE become more competitive in Hit@$k$, but diffusion consistently maintains higher Hit@$k$ across all sweep budgets, while achieving comparable SNR\,ratio@$k$.

\paragraph{Diffusion vs.\ reference priors}
Diffusion models substantially outperform reference priors across all sweep budgets. AoA-based, uniform, and random priors exhibit low Hit@1 values (below $0.16$) and slower improvement with increasing $k$, whereas diffusion achieves consistently higher Hit@$k$ and smoother scaling behavior, with comparable SNR\,ratio@$k$.

\subsection{Diffusion Ablations and Sampling Comparison}

\begin{figure}[t]
\centering
\subfloat[Hit@$k$, large models]{%
    \includegraphics[width=0.48\linewidth]{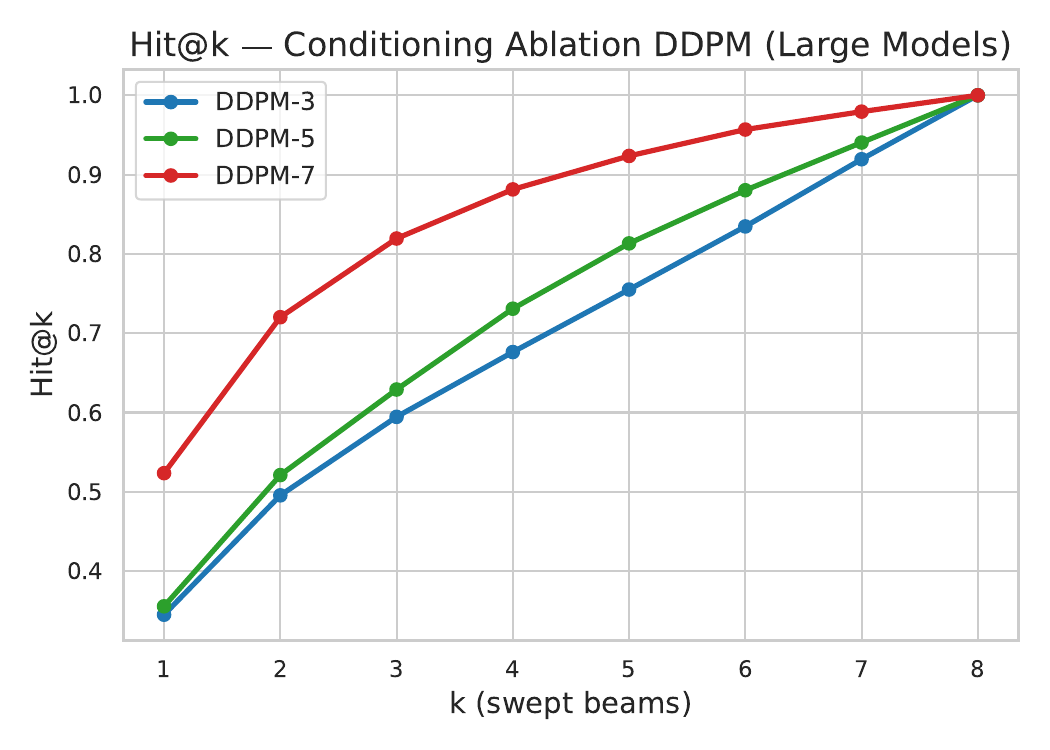}}
\hfill
\subfloat[Hit@$k$, small models]{%
    \includegraphics[width=0.48\linewidth]{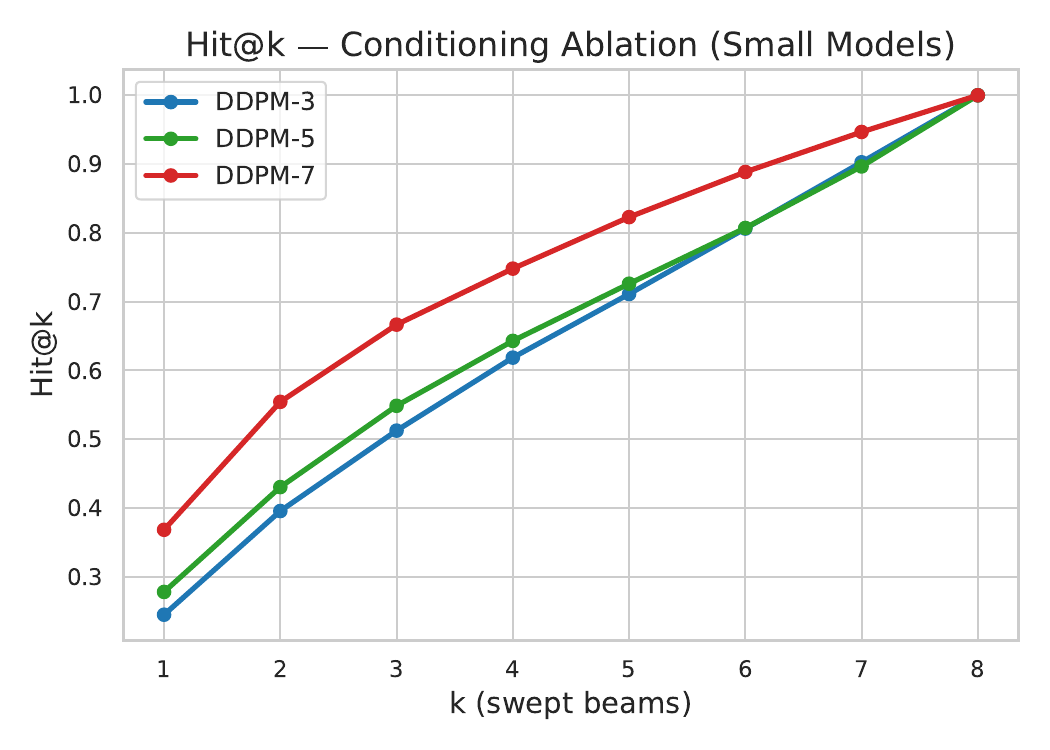}}
\vspace{0.15cm}
\subfloat[SNR, large models]{%
    \includegraphics[width=0.48\linewidth]{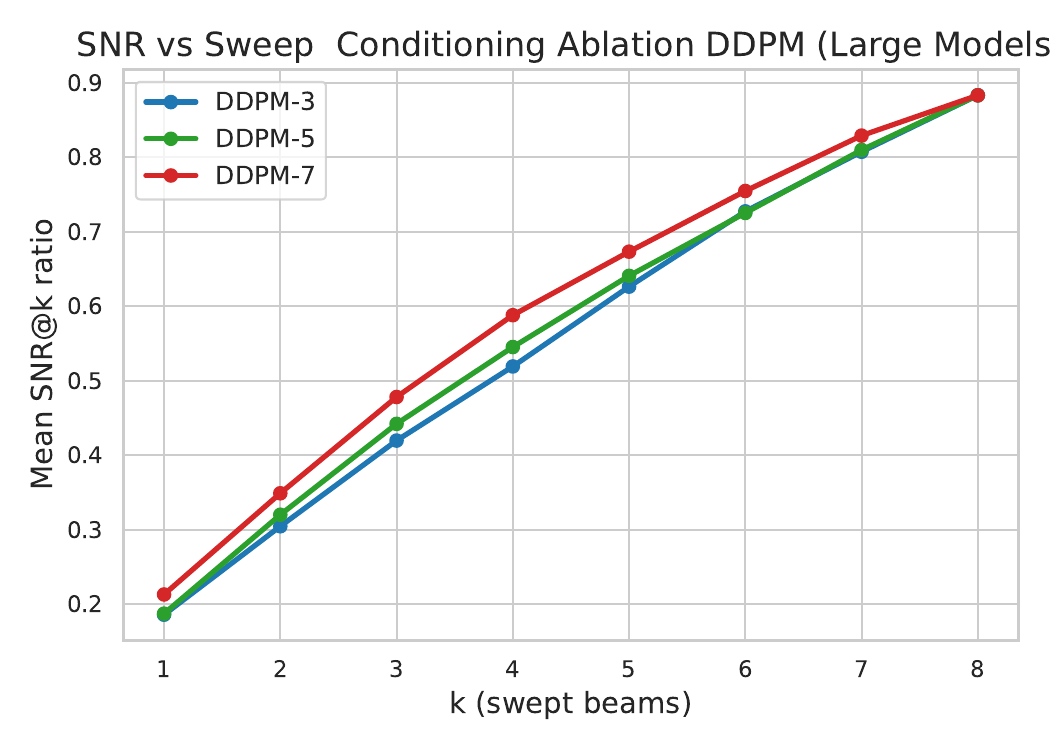}}
\hfill
\subfloat[SNR, small models]{%
    \includegraphics[width=0.48\linewidth]{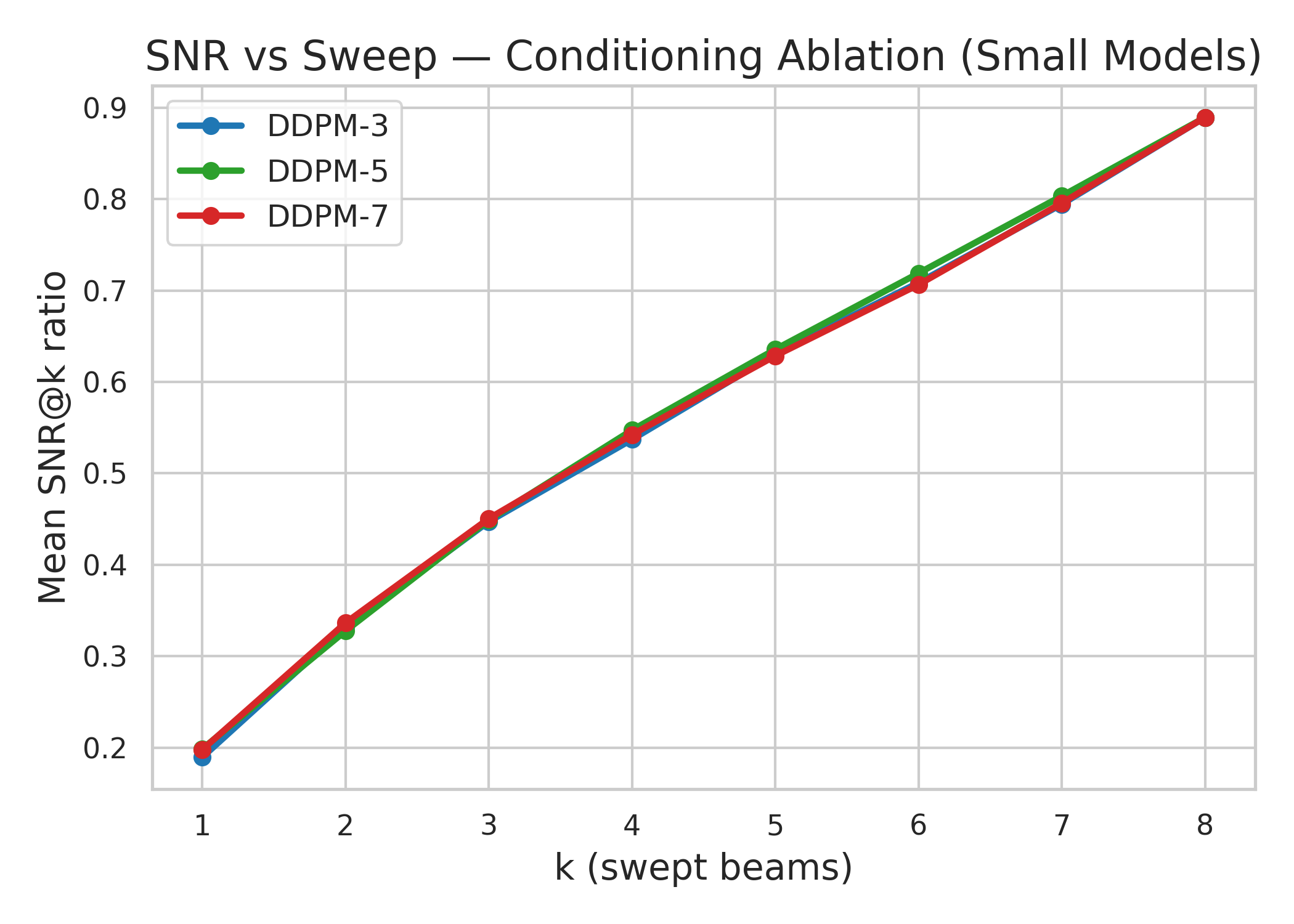}}
\vspace{0.15cm}
\subfloat[DDPM vs.\ DDIM vs. UNet Hit@$k$]{%
    \includegraphics[width=0.48\linewidth]{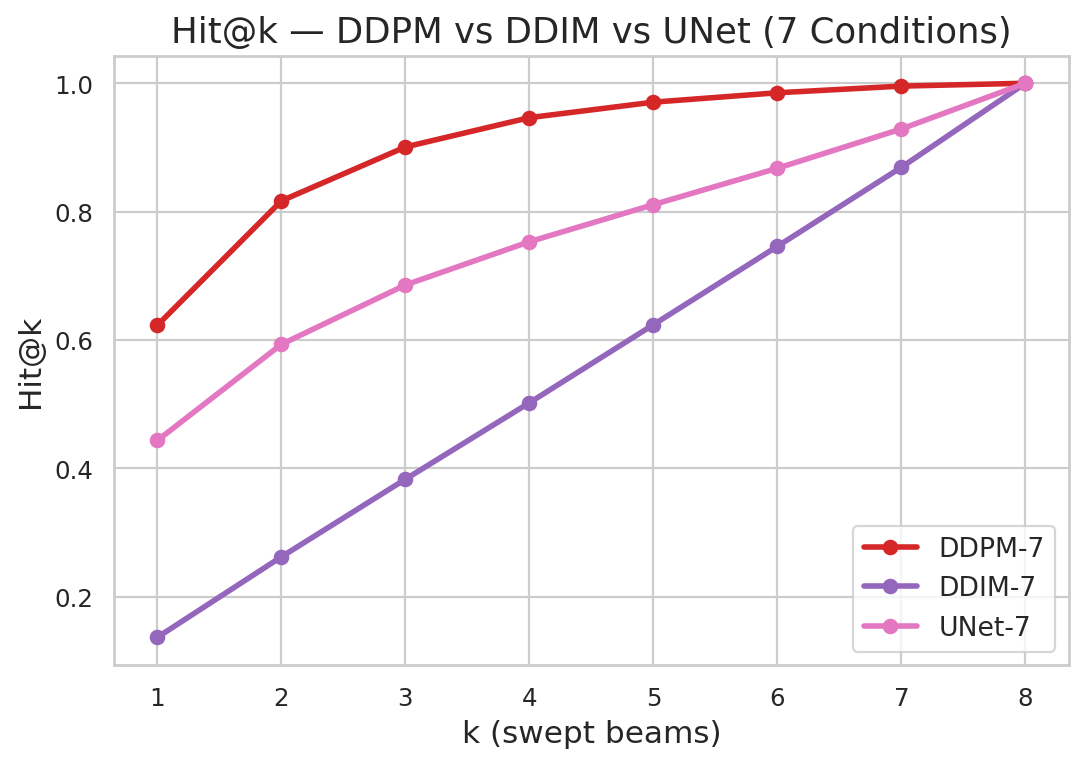}}
\hfill
\subfloat[DDPM vs.\ DDIM vs. UNet SNR]{%
    \includegraphics[width=0.48\linewidth]{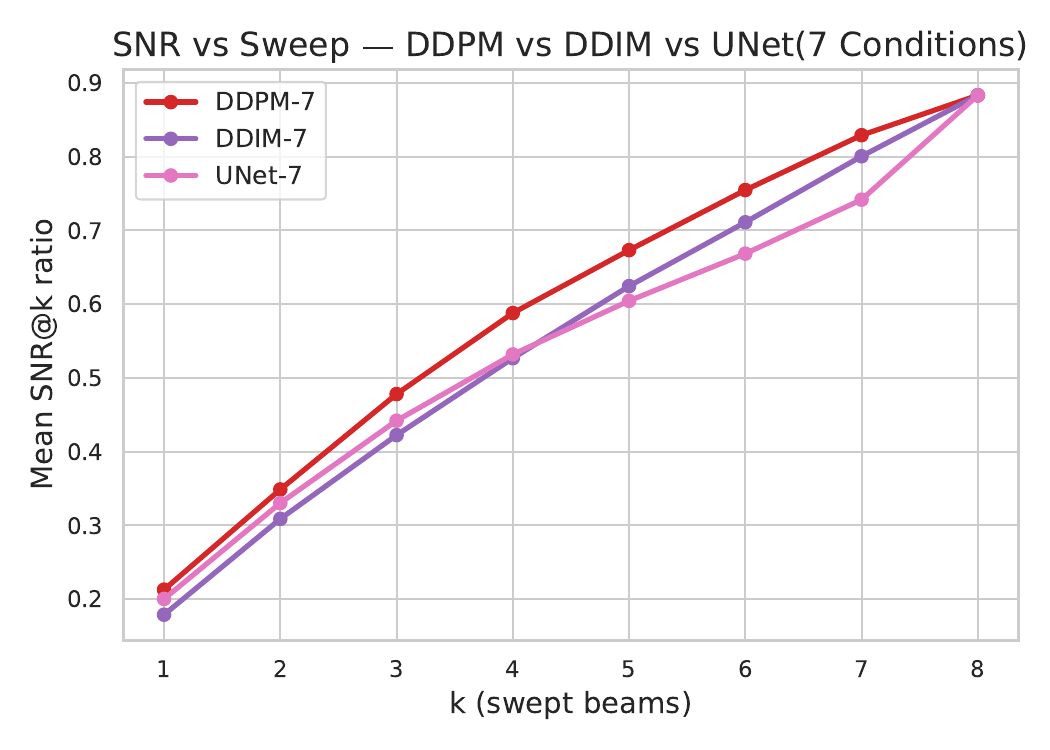}}

\caption{Diffusion ablations across conditioning dimensionality, model size, and sampling strategy. Increasing conditioning from 3D to 5D and 7D consistently improves Hit@$k$, while SNR@$k$ remains largely stable across feature sets for both small and large denoisers. With 7D conditioning, DDPM achieves the highest Hit@$k$, followed by UNet-based DDPM and DDIM, while SNR@$k$ differs only marginally across methods.}

\label{fig:diffusion_ablations}
\end{figure}

We examine the impact of conditioning dimensionality, model capacity, and sampling strategy on diffusion-based beam alignment performance, as illustrated in Fig.~\ref{fig:diffusion_ablations}.

\paragraph{Effect of conditioning dimensionality}
For both small and large denoisers, performance improves consistently as the conditioning dimensionality increases from 3D to 5D and further to 7D. For the large model, Hit@5 increases from approximately $0.76$ (3D) to $0.81$ (5D) and $0.92$ (7D), while the small model exhibits a corresponding increase from $0.70$ to $0.72$ and $0.82$. In contrast, SNR\,ratio@$k$ shows only marginal variation across conditioning sets. These results suggest that incorporating additional geometric and propagation aware features such as distance, LOS or NLOS indicators, and strongest path AoA and AoD primarily enhances beam ranking accuracy, with no impact on the received SNR.

\paragraph{Effect of model capacity}
For a fixed conditioning set, small and large denoisers yield closely aligned Hit@$k$ and SNR\,ratio@$k$ trends. Under 7D conditioning, the DDPM model achieves the strongest overall performance in both settings. Specifically, the large DDPM-7 model attains Hit@5 of $0.92$ and SNR@5 of $0.69$, compared to Hit@5 of $0.82$ and SNR@5 of $0.62$ for the small model. This improvement comes at the cost of increased model size, with $1.39$M parameters for the large model versus $0.236$M parameters for the small model. These results suggest that performance is primarily driven by the richness of side information, with model capacity providing only a secondary benefit, indicating that relatively compact diffusion models are sufficient for this task.

\paragraph{DDPM vs.\ DDIM vs.\ UNet}
The bottom row of Fig.~\ref{fig:diffusion_ablations} compares sampling strategies and denoiser architectures under 7D conditioning. When comparing sampling methods with a fixed denoiser, DDPM sampling with $500$ denoising steps achieves a Hit@$5$ of $0.98$, but incurs significantly higher inference cost, with approximately $0.48$ ms/user latency and $6.4\times10^{-2}$ J/user energy consumption. In contrast, DDIM sampling with $50$ steps reduces latency and energy by nearly
an order of magnitude ($\sim 0.05$~ms/user and $6.9\times10^{-3}$~J/user),
at the expense of lower Hit@$k$ at small sweep budgets, achieving Hit@$5$ of $0.61$.

When comparing denoiser architectures under DDPM sampling, the baseline DDPM denoiser with $1.39$M parameters outperforms the UNet-based DDPM model in Hit@$k$ (Hit@$5$ of $0.98$ versus $0.80$), while also achieving lower latency and energy consumption ($\sim0.48$ ms/user and $6.2\times10^{-2}$ J/user versus $\sim0.54$ ms/user and $7.7\times10^{-2}$ J/user), despite the UNet-based model having a larger parameter count of $1.85$M. Across all variants, SNR@$k$ remains largely comparable.

\subsection{Discussion}

Diffusion-based beam priors provide the best tradeoff between beam ranking accuracy and SNR preservation at small sweep budgets, which are the most relevant operating points for low-overhead and energy-efficient beam alignment. In this regime, diffusion significantly outperforms deterministic, generative, and heuristic baselines in Hit@$k$, while maintaining comparable SNR\,ratio@$k$.

Ablation results indicate that the observed performance gains stem mainly from richer conditioning information rather than increased model capacity. Comparisons across sampling strategies further expose a clear accuracy–complexity trade-off: DDPM achieves the highest small-$k$ Hit@$K$ performance at the cost of higher latency and energy consumption, whereas DDIM substantially reduces inference cost with only moderate accuracy degradation at small sweep budgets.

\section{Conclusion}

This work establishes conditional diffusion-based beam priors as a powerful and practical tool for resource-efficient beam alignment in directional mmWave and THz systems. By learning uncertainty-aware beam distributions from compact side information, the proposed framework enables accurate alignment with significantly reduced beam sweeping overhead, while preserving received SNR. Extensive evaluations and ablation studies demonstrate that diffusion models deliver state-of-the-art ranking performance at the most resource-constrained operating points, and offer flexible accuracy–latency–energy tradeoffs through controllable sampling strategies. These results position diffusion-based beam alignment as a strong candidate for next-generation low-overhead wireless systems.

\bibliographystyle{IEEEtran}
\bibliography{references}


\end{document}